\documentclass[aps,prd]{revtex4}
\usepackage{graphicx}

\newcommand\bef{\begin{figure}}
\newcommand\eef[1]{\label{fg:#1}\end{figure}}
\newcommand\beq{\begin{equation}}
\newcommand\eeq[1]{\label{#1}\end{equation}}
\newcommand\beqa{\begin{eqnarray}}
\newcommand\eeqa[1]{\label{#1}\end{eqnarray}}
\newcommand\bet{\begin{table}}
\newcommand\eet[1]{\label{tb:#1}\end{table}}

\newcommand\fgn[1]{Figure \ref{fg:#1}}
\newcommand\eqn[1]{eq.\ (\ref{#1})}

\newcommand\cl[1]{#1\%{\rm\ CrI\/}}
\newcommand{\D}{{\cal D}}
\newcommand\etc{{\sl etc.\/}}
\newcommand\etal{{\sl et al.\/}}
\newcommand\ie{{\sl i.e.\/}}
\newcommand{\I}{{\cal I\/}}
\newcommand{\ifr}{{\rm IFR\/}}
\newcommand{\sfr}{{\rm SFR\/}}
\newcommand{\T}{{\cal T}}

\begin{document}

\author{Sourendu Gupta}
\affiliation{Department of Theoretical Physics, Tata Institute of Fundamental
 Research,\\ Homi Bhabha Road, Mumbai 400005, India.}
\title{Inferring epidemic parameters for COVID-19 from fatality counts in Mumbai}
\begin{abstract}
Epidemic parameters are estimated through Bayesian inference using
the daily fatality counts in Mumbai during the period from March 31 to
April 14.  A doubling time of 5.5 days (median with $[4.6,6.9]$ days
\cl{95}) is observed. In the SEIR model this gives the basic reproduction
rate $R_0$ of 3.4 (median with $[2.4,4.8]$ \cl{95}). Using as input the
infection fatality rate and the interval between infection and death, the
number of infections in Mumbai is inferred. It is found that the ratio of
the number of test positives to the total infections is 0.13\% (median),
implying that tests are currently finding 1 out of 750 cases of infection.
After correcting for different testing rates, this result is compatible
with a measurement of the ratio made recently via serological testing
in the USA. From the estimates of the number of infections we infer that
the first COVID-19 cases were seeded in Mumbai between late December 2019
and early February 2020. provided the doubling times remained unchanged
since then. We remark on some public health implications if the rate of
growth cannot be controlled in about a week.
\\ \bigskip
TIFR/TH/20-11\\
\end{abstract}
\maketitle

\section{Introduction}

In many countries across the world, there is an initial bottleneck in
testing for COVID-19. This is due to a combination of two factors, first
the availability of test kits, and second, of trained laboratory personnel
to perform the tests. When the number of tests per million of population
is limited, the exponential growth in infections at the beginning of an
epidemic makes it impossible to estimate the prevalence of infections in
any meaningful way. One aim of this paper is to establish and validate a
Bayesian statistical inference procedure to make a probabilistic estimate
of the infected population from the statistics of fatalities. This is
done using data for Mumbai, which is one of the major COVID-19 epidemic
hot spots in India.

The counts of fatalities due to COVID-19 also have errors. The most
important is the fraction of mis-diagnosis. At the current time, it is
unlikely that a patient who dies of severe acute respiratory illness
(SARI) will be misdiagnosed. We know that the Municipal Corporation of
Greater Mumbai (MCGM) tries to track, and report on, every such case,
even if the diagnosis comes a few days after the fatality. Another source
of systematic error in the count of fatalities may come from a part of
the population with limited access to health care. This is not a trivial
source of error, and corporations have to check their death records very
thoroughly to make sure that no case is missed. In New York City, where
the number of daily fatalities attributed to COVID-19 has been more than
100 since late March \cite{nychealth} the local government is wary of
undercounts \cite{reuters}. Similar undercounts have been reported from
several European countries \cite{forbes}, and a correction by 100--200\%
was found to be necessary. A retrospective correction in the record of
fatalities is being carried out in Wuhan. Nevertheless, fatality counts
may be in error by a factor 2--3, whereas testing could potentially
have errors which are one or two orders of magnitude higher.

In view of this, it makes sense to estimate epidemic parameters from the
daily count of fatalities, $D(t)$, rather than less uncertain statistics
such as the daily count of test positives $T_+(t)$.
This is the primary aim of this paper. Daily fatality counts in Mumbai
are currently less than 20. This is about one part in a million of the
population, and an order of magnitude below peak rates in cities of
comparable size elsewhere in the world. The simplifying assumption that
the epidemic is still in the stage of exponential growth is then viable.
The exponential growth rate can then be reliably extracted from $D(t)$
without using any detailed model. Detailed analysis of the data also
allows validation of this assumption, as we discuss in the results
section of this paper. 

Two more parameters, the infection fatality rate, $\ifr$, and the interval
from infection to death, $T$, relate the number of infections, $I(t)$,
and $D(t)$.  Since data on $I(t)$ in Mumbai are currently unreliable,
these two parameters are taken from the literature \cite{ifr} to predict
a credible interval for $I(t)$. Alternative estimates of $I(t)$ would,
of course, constrain these parameters.  First reports of such alternative
estimates are now available from other locations \cite{seroprev}.  We find
that the ratio $\Pi=T_+(t)/I(t)$ is small, and has been almost independent
of time in Mumbai. This ratio has also been reported in \cite{seroprev}.
It is checked that the estimate of $\Pi$ made here is compatible with
that, provided that the different testing rates in different countries
are taken into account.

These statistical inferences lead directly to estimates of the date at
which the initial seeding of COVID-19 took place. The further assumption
which goes into this estimate is that the growth rate has remained
unchanged. Another caveat should be kept in mind. Due to the statistical
nature of infections, it is possible that multiple seedings occurred,
and that the date obtained by the procedure followed here gives an average
over such statistics. Stochastic models would refine our understanding of
the early growth process. Genomic studies of multiple samples of the virus
\cite{yadav}
in Mumbai would be independent data which could constrain such models.

\bef
\begin{center}
 \includegraphics[scale=0.5]{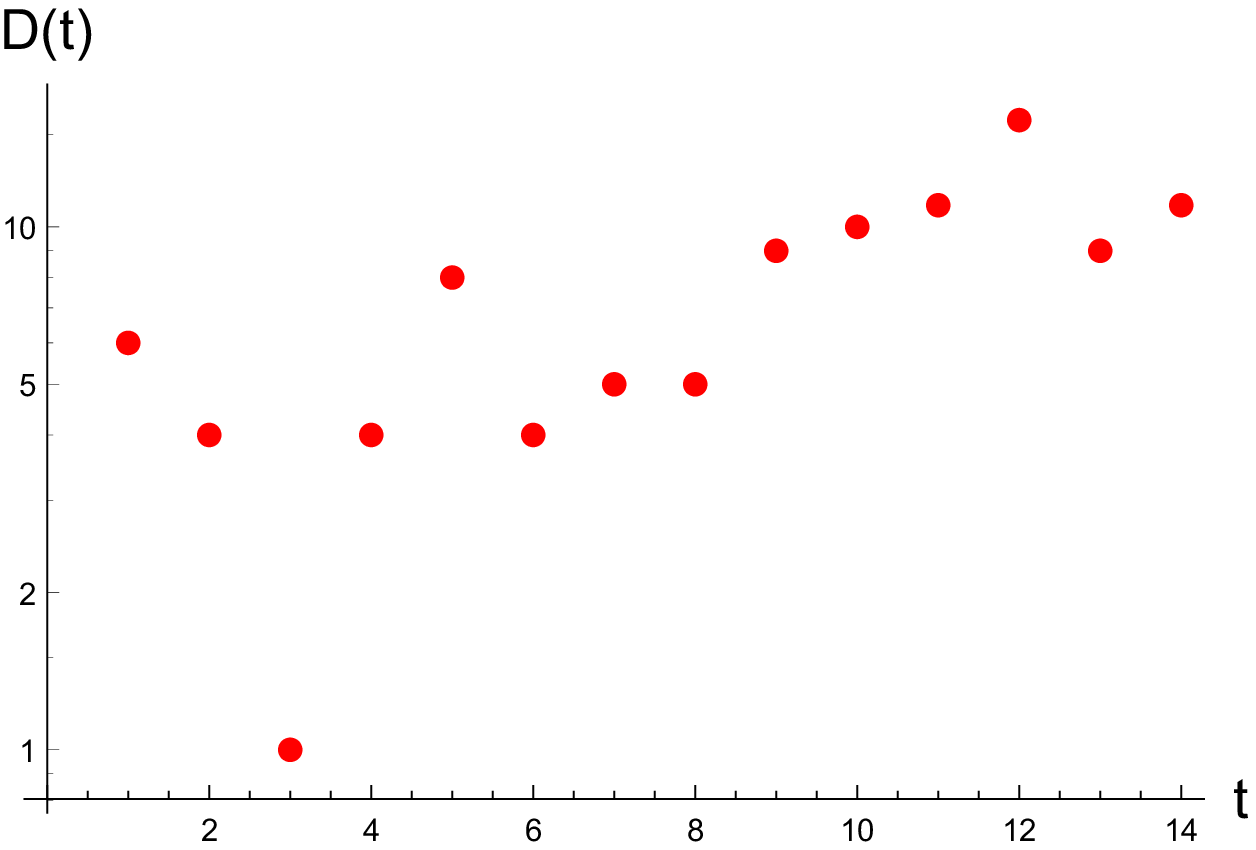}
\end{center}
\caption{The number of daily fatalities due to COVID-19 in Mumbai on
 every day in April up to April 14. More recent data has not been 
 analyzed together with this, because retrospective corrections may
 be applied for those. Notice the log scale on the $y$-axis.}
\eef{deaths}

MCGM releases a daily count of fatalities due to COVID-19, and performs
retrospective corrections due to late receipt of test results or other
post-mortem analysis. Due to this, the count of fatalities may take a
few days to settle down. Changes in municipal priorities may also affect
resources assigned to various tasks, resulting in systematic changes
in the statistical character of data streams. Any analysis of epidemic
parameters must allow for checks of such confounding factors. Other
data streams, such as the municipal records of deaths, are harder to
access. However, such records will need to be subjected to statistical
analysis later to ascertain whether there were possible cryptic fatalities
due to COVID-19. The methods used here can also be utilized to analyze
the results of such an appraisal.

\section{Method}

\subsection{Data used}

Daily counts released by MCGM have been archived \cite{prahlad}. Here
the summed and anonymous data on fatalities is analyzed for the two-week
period between May 30 and April 14. The basic data on counts is displayed
in \fgn{deaths}. No clear trend can be distinguished in the first week,
but a clear exponential growth, characteristic of the early stages of
epidemics, can be seen in the second week.

\subsection{Mathematical formulation: estimation of doubling time}

During the early stages of an epidemic the number of fatalities is expected
to grow exponentially, doubling in every $\tau$ days,
\beq
 D(t)=\D_0 2^{t/\tau}.
\eeq{doubling}
The parameter $\D_0$ is an estimate of the number of fatalities at the
time when the counting starts. Both $\tau$ and $\D_0$ will be treated
as parameters to
be determined from the time series $D(t)$ using a standard Bayesian
inference process.

Given the set of parameters $\pi=\{\tau,\D_0\}$, and the set of data
to be fitted $\{D_1, D_2,\cdots\}$, which are the fatalities on days 1,
2, \etc, and the errors $\{\sigma_1,\sigma_2,\cdots\}$, the probability
distribution function (PDF) of $D_i$ given $\pi$ is defined as
\beq
   P(D_1,D_2,\cdots|\pi) = \exp\left[-\sum_t 
       \frac{[D_t-D(t)]^2}{\sigma_t^2} \right]
\eeq{chisq}
where $D(t)$ is the model of \eqn{doubling}, which depends on $\pi$.
With the present disease surveillance regime, significant miscounts of
$D_t$ would come only from small clusters of unnoticed infections, which
would grow at roughly the same rate as the rest of the population. So
the error would be a fraction of $D_t$.  We will take $\sigma_t=f D_t$,
with two scenarios of $f$. In scenario L, the low error scenario, we
take $f=0.1$, \ie, assume 10\% errors on $D_t$.  In scenario M, the
medium error scenario, we double the errors and take $f=0.2$. Bayes'
theorem can be used to write the posterior PDF for $\pi$,
\beq
   P(\pi|D_1,D_2,\cdots) \propto P(D_1,D_2,\cdots|\pi) P_0(\pi)
\eeq{bayes}
where $P_0(\pi)$ is called the prior PDF of the parameters The constant
of proportionality, which is the prior PDF of the data, is trivial, since
it does not depend on $\pi$, and serves only to normalize the posterior
PDF. Maximizing the posterior PDF gives the best values of $\pi$. We
will also quote the 95\% credible intervals (\cl{95}) obtained from the
posterior distribution.

The prior PDF is taken to be the product of two independent Gamma
distributions, one for each of the parameters, $p_0(\D_0) = \Gamma(\D_0;
k_0,\theta_0)$ and $p_0(\tau) = \Gamma(\tau;k,\theta)$.  We take the
meta-parameter $\theta=\theta_0=10$, so that we don't suffer from
overconfidence in priors \cite{lahiri}.  The parameters $k$ and $k_0$
are chosen to give the most probable prior values of $\tau=10$ days
and $\D_0=5$. It was checked that different choices of prior parameters
gave results which were insensitive to the choices.

\subsection{Mathematical formulation: estimation of number of infections}

$I_0$, the number of infections on day zero, can be inferred from $\D_0$
by a straightforward argument \cite{previous}.  First, the infection
fatality rate, $\ifr$, gives a statistical estimate of the number of
infections, $\I$, which resulted in the observed fatalities on day
zero. This is simply $\I=\D_0/\ifr$.  The value of $\ifr$ could change
from country to country because of differing age and gender structure of
populations, or due to prevailing health conditions. However, a larger
source of uncertainty is in the fraction of asymptomatic cases. As a
result, $\ifr$ estimates vary from over 1\% \cite{russell} to under 0.5\%
\cite{cebm}. Based on an complete sampling of a well-defined population,
it has been estimated that the age and gender averaged $\ifr$ is 0.657\%
\cite{ifr}. This is the estimate that we use here. A recent preprint
\cite{goli} updates the work in \cite{previous} by correcting this value
using Indian census data to obtain a value which is approximately two
thirds of this. Using this number would boost our median estimates by
about 50\%. Since the \cl{95} are much larger, we do not incorporate
this correction.

From $\I$ it is possible to construct $I_0$ using the previous estimate
of $\tau$ to write $I_0=2^{T/\tau}\I$. Also, $T=t_1+t_2$ where $t_1$
is the incubation period, \ie, the interval between infection and the
occurrence of symptoms, and $t_2$ is the interval between the occurrence
of symptoms and death. Several independent estimates agree that the
median value of $t_1$ is around 5 days \cite{lauer,li}. For $t_2$ we
use the mean of 18.8 days and coefficient of variation of 0.25 \cite{ifr}.

Putting these together we have
\beq
   I_0 = \D_0\left[\frac1{\ifr}\times 2^{T/\tau}\right],
    \qquad{\rm with}\qquad T=t_1+t_2,
\eeq{ioest}
and the distributions of the three parameters are taken to be
$\Gamma(\ifr;6.3,0.0011)$, $\Gamma(t_1;2.06,2.93)$ and
$\Gamma(t_2;16,1.175)$. The amplification factor, within square brackets,
has all the uncertainties associated with these inputs, unlike the
parameters $\D_0$ and $\tau$. Nevertheless, $I_0$ is of high interest
for public health reasons, so we will present estimates.

\section{Results}

\bef
\begin{center}
\includegraphics[scale=0.5]{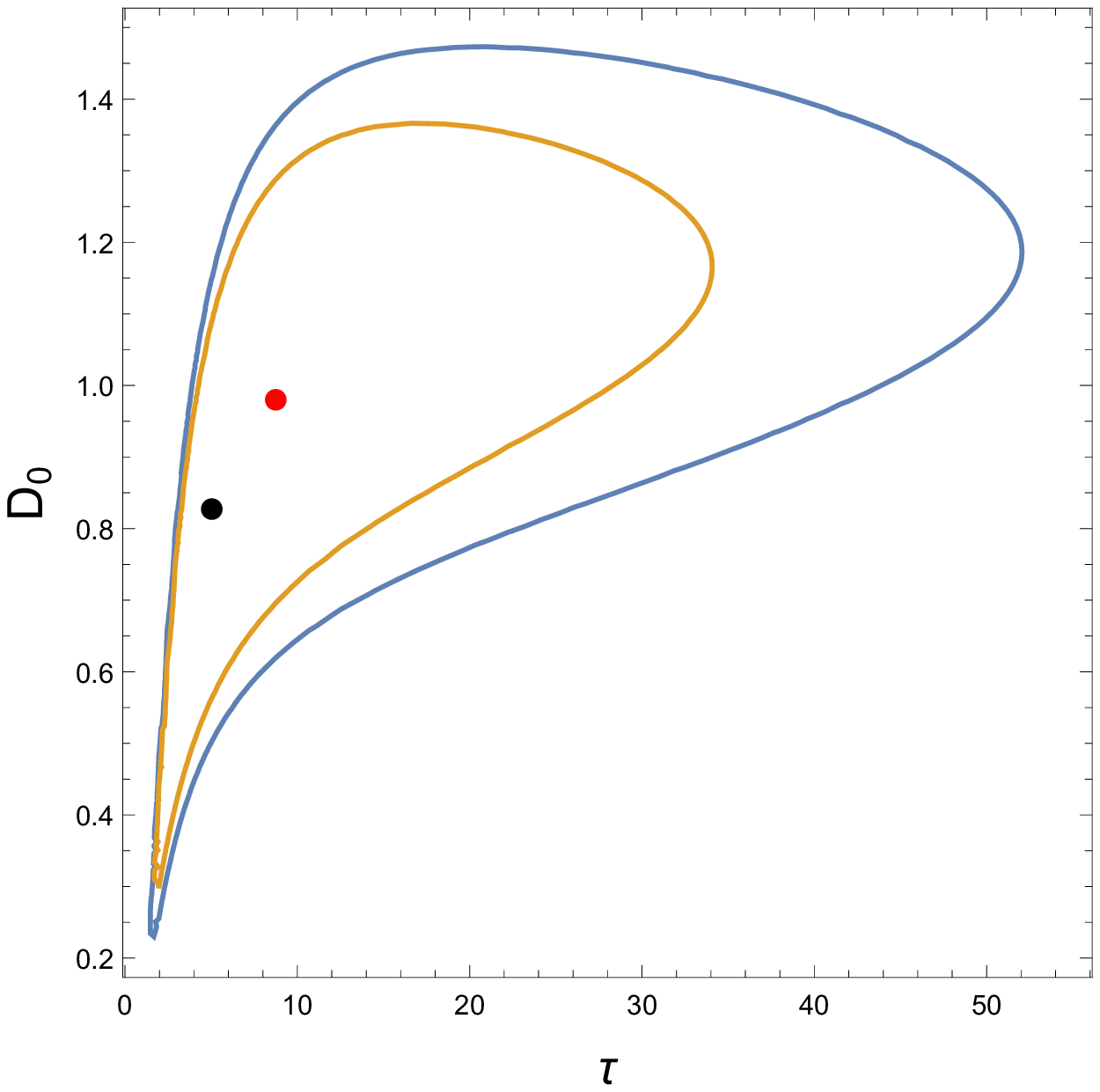}
\includegraphics[scale=0.5]{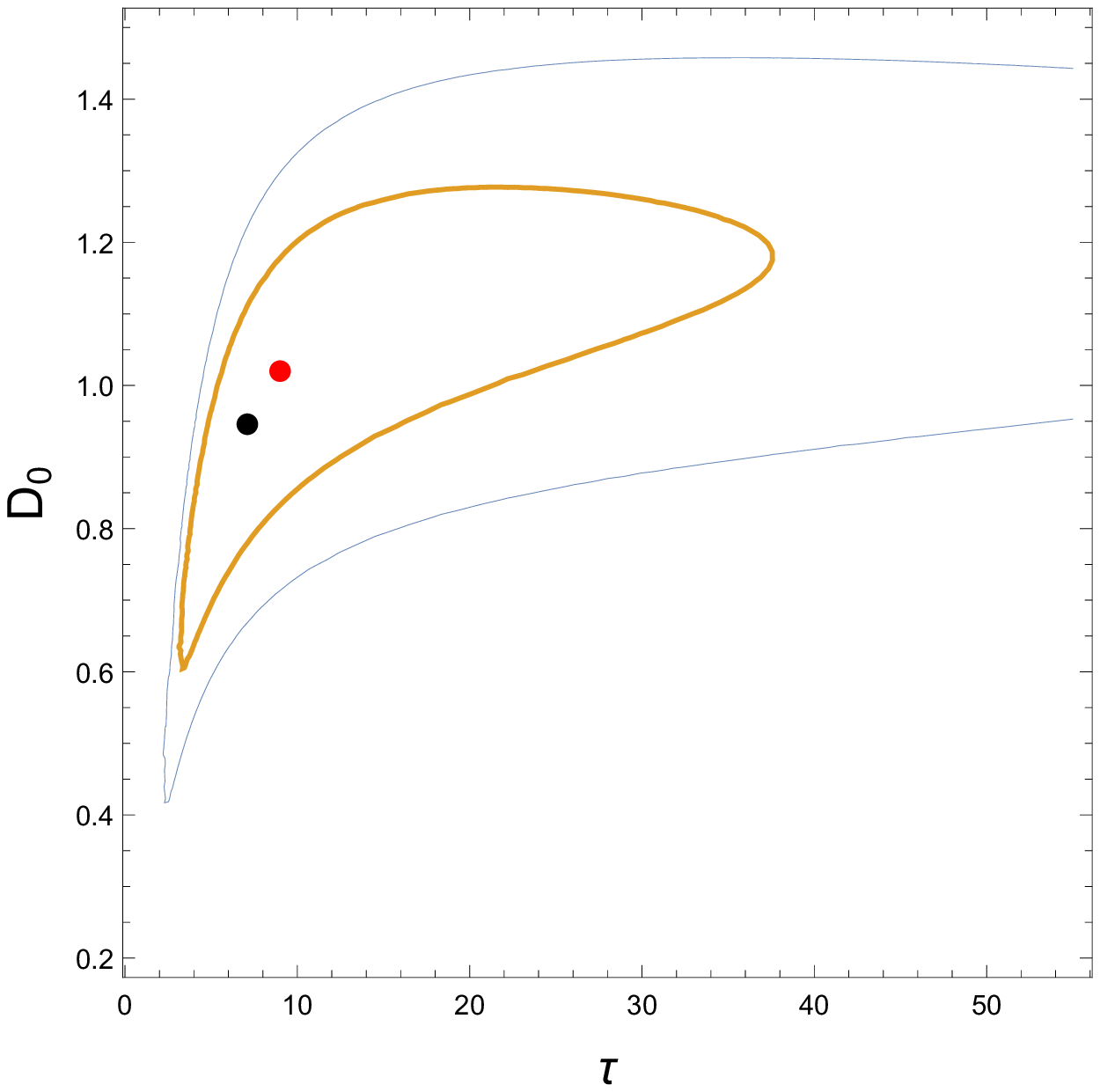}
\end{center}
\caption{The best fit values of $\tau$ and $\D_0$ assuming 20\% errors on
 the number of fatalities using the data between March 31 and April 4. The
 panel on the left is for Scenario M and that on the right for Scenario L.
 The best fit (modal values) given in \eqn{modelacentral} are denoted by
 the black dots, and the medians by the red dots.
 The inner curve encloses the \cl{95}, and the outer curve the \cl{99} (for
 Scenario M) or the \cl{97} (for Scenario L) for which the \cl{99} curve is
 very far away on this scale.}
\eef{firstpart}

\subsection{March 31 to April 4}

The first analysis was performed with the short
time series for fatalities between March 31 and April 4. The most likely
model that emerges out of the data is
\beqa
 \nonumber {\rm Scenario\ M:\/}&&\quad
 \tau = 5.0{\rm\ days},\quad
 \D_0 = 0.83, \quad{\rm which\ implies}\quad
 I_0 = 17000\;[2000,350000]\;\cl{95}.\\
 {\rm Scenario\ L:\/}&&\quad
 \tau = 7.1{\rm\ days},\quad
 \D_0 = 0.95, \quad{\rm which\ implies}\quad
 I_0 = \phantom{0}4600\;[\phantom{0}900,\phantom{0}42000]\;\cl{95}.
\eeqa{modelacentral}
The joint 95\% and \cl{99} on the model parameters are shown in
\fgn{firstpart}.  Note that the posterior distribution of the model
parameters is strongly non-Gaussian. From the posterior PDF for $\tau$, marginalized over $\D_0$
in scenario M, we find $\tau$ to be 8.75 days (median values) and $[3.2,30]$ days \cl{95},
giving a significant probability of very long doubling times. From the PDF for $\D_0$,
marginalized over $\tau$, we find a median of 0.98 and $[0.6,1.3]$
\cl{95}. One sees in \fgn{firstpart} that there is a broad tail to the
posterior PDF which allows slightly larger values of $\D_0$ to compensate
for a much slower epidemic growth. Longer time series are needed to rule this
out, as will be seen later. 

In Scenario L, after marginalizing over $\D_0$,
$\tau$ is found to be 9.0 days (median) with $[4.8,27]$ days \cl{95}. For $\D_0$ 
it was seen that $\D_0=1.02$ and $[0.8,1.2]$ \cl{95}. The wide
difference between the \cl{95} and the \cl{99} curves in scenario L is
an indication that 10\% Gaussian errors in scenario L are unable to give
a good description of the jitter in the time series. In view of this, scenario L is not
used in subsequent analyses.

\bef
\begin{center}
\includegraphics[scale=0.5]{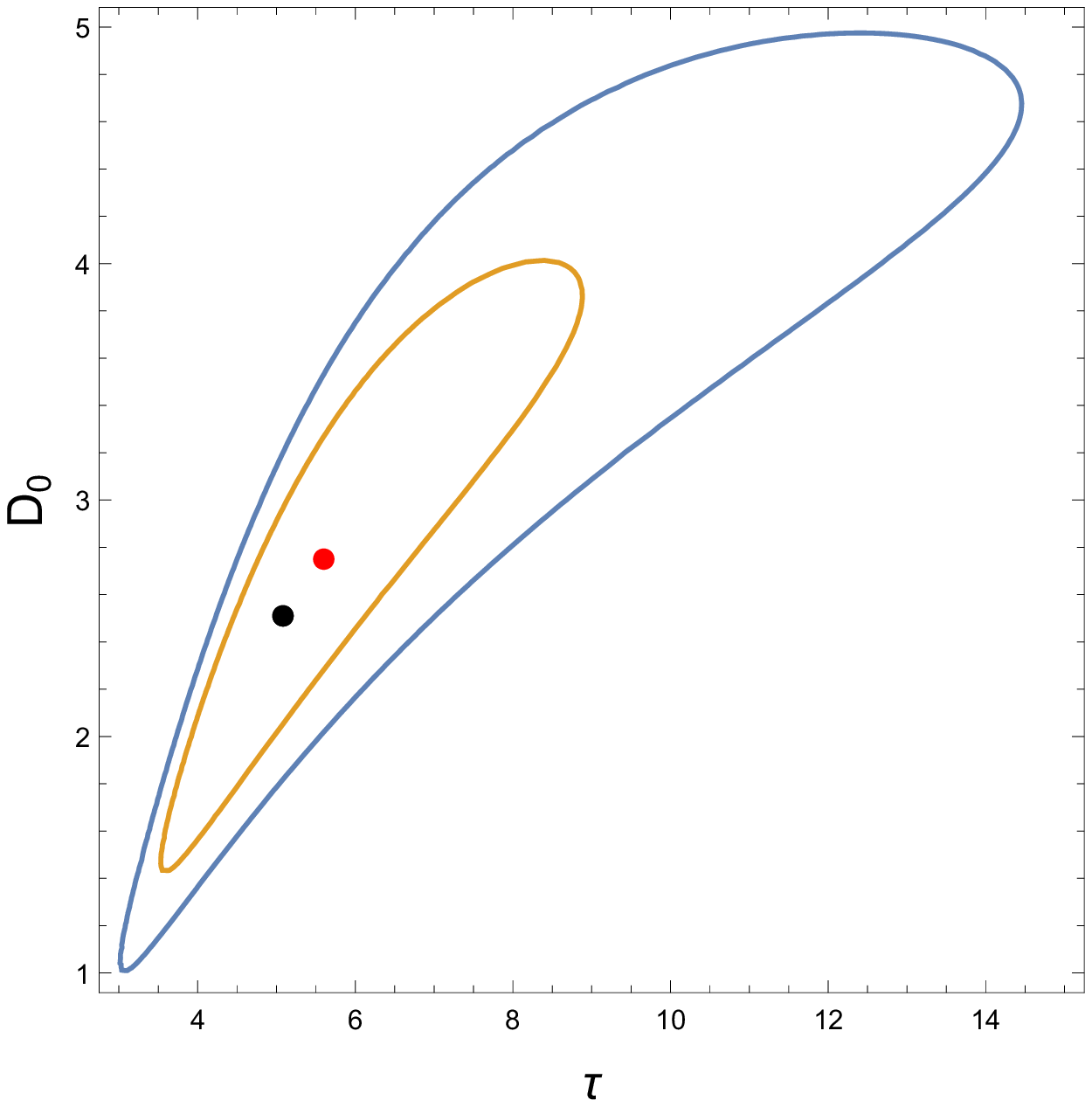}
\end{center}
\caption{The best fit values of $\tau$ and $\D_0$ (day zero chosen to
 be April 2) in Scenario M using data between April 6 and 14. 
 The best fit (modal values) is denoted by
 the black dot, and the median by the red dot.
 The inner curve encloses the \cl{95}, and the outer one the \cl{99}.}
\eef{secondpart}

\bef
\begin{center}
\includegraphics[scale=0.5]{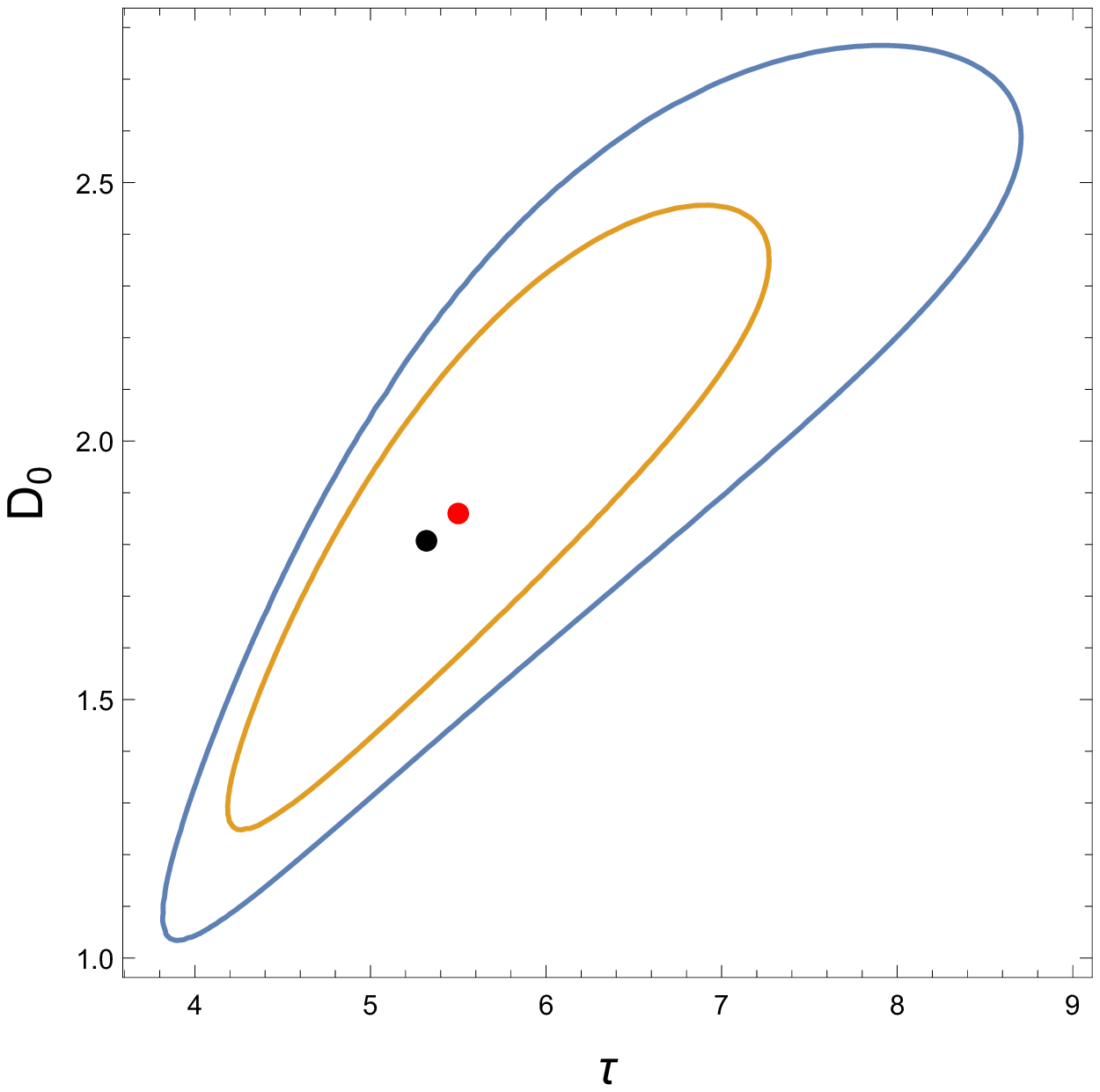}
\end{center}
\caption{The best fit values of $\tau$ and $\D_0$ (on March 30) using
 the data set between March 31 and April 14, corrected for the change in
 intensity of surveillance soon after April 2, as described in the text.
 The best fit (modal values) is denoted by
 the black dot, and the median by the red dot.
 The inner curve encloses the \cl{95}, and the outer one the \cl{99}.}
\eef{fullset}

\subsection{April 6 to April 14}

The second part of the analysis used the data for the eight days between
April 6 and 14, with day zero to be April 2.  So $\D_0$ and $I_0$ in this
context are estimates of the inferred number of infected cases in Mumbai
on April 2.  Note that this time series is almost 50\% longer than the
first, and therefore regulates some of the peculiarities seen earlier.
The most likely model parameters in Scenario M are
\beq
 \tau = 5.1{\rm\ days},\quad
 \D_0 = 2.5, \quad{\rm which\ implies}\quad
 I_0 = 44000\;[5500,960000]\;\cl{95}.
\eeq{modelbcentral}
The 95\% and \cl{99} on the joint distribution of model parameters are
shown in \fgn{secondpart}. Although the contours are not as distorted
as in \fgn{firstpart}, they are clearly not ellipses. So the posterior
PDF is distinctly non-Gaussian. After marginalizing over $\D_0$, we find
$\tau$ is 5.6 days (median) and $[4.1,8.4]$ days \cl{95}.  This \cl{95}
is fully contained in the \cl{95} of $\tau$ for the earlier data set,
showing that the estimates are totally compatible.  For $\D_0$ we find,
after marginalizing the posterior PDF over $\tau$, 2.75 (median) and
$[1.8,4]$ \cl{95}.

Given the range of $\tau$ quoted above, in 3 days between March 31 and
April 2, one expects $\D_0$ on May 31 to increase by a factor $[1.3,1.7]$
\cl{95}, and become $[1.0,2.0]$ \cl{95}. While this interval overlaps
with that quoted above, the probability that the two estimates are
statistically indistinguishable is about 3.8\%. The reason for this
discrepancy can be traced to a change in the surveillance policy
instituted on April 2 \cite{news}. The increased vigilance after this
date must be corrected for in order to compare the two data sets. One can
account for this by changing the overall normalization of the earlier
data set by a factor of 1.7. This then increases the probability of
agreement between the two analyses by an order of magnitude.

If one restricts the time series between April 6 and 12, one sees ``by
eye'' that there could be rapid exponential growth. For this specially
selected data set, we find that the most probable $\tau$ is 2.8 days.
There is only a 20\% probability that $\tau$ is as large as 4 days or
more.  These alarming results are biased.  It is important to remember
that infection spreading is a stochastic process, and there are always
short runs of low or high growth.

\subsection{The full set March 31 to April 14}

Third, the analysis of the full data set is presented. A reporting
discrepancy between the series up to April 4 and the later part was
discussed in the previous section. We applied the correction developed
there, and rounded $D(t)$ to the nearest integer. 
This corrected time series gave
\beq
 \tau = 5.3{\rm\ days},\quad
 \D_0 = 1.8, \quad{\rm which\ implies}\quad
 I_0 = 27000\;[3500,520000]\;\cl{95}.
\eeq{model}
The \cl{95} of the joint distribution is shown in \fgn{fullset}. The PDF
for $\tau$ marginalized over $\D_0$ gives $\tau=5.5$ days (median) with
$[4.6,6.9]$ days \cl{95}, and that for $\D_0$ marginalized over $\tau$
gives $\D_0=1.86$ (median) with $[1.45,2.35]$ \c{95}. These are our
primary results.

\bef
\begin{center}
\includegraphics[scale=0.5]{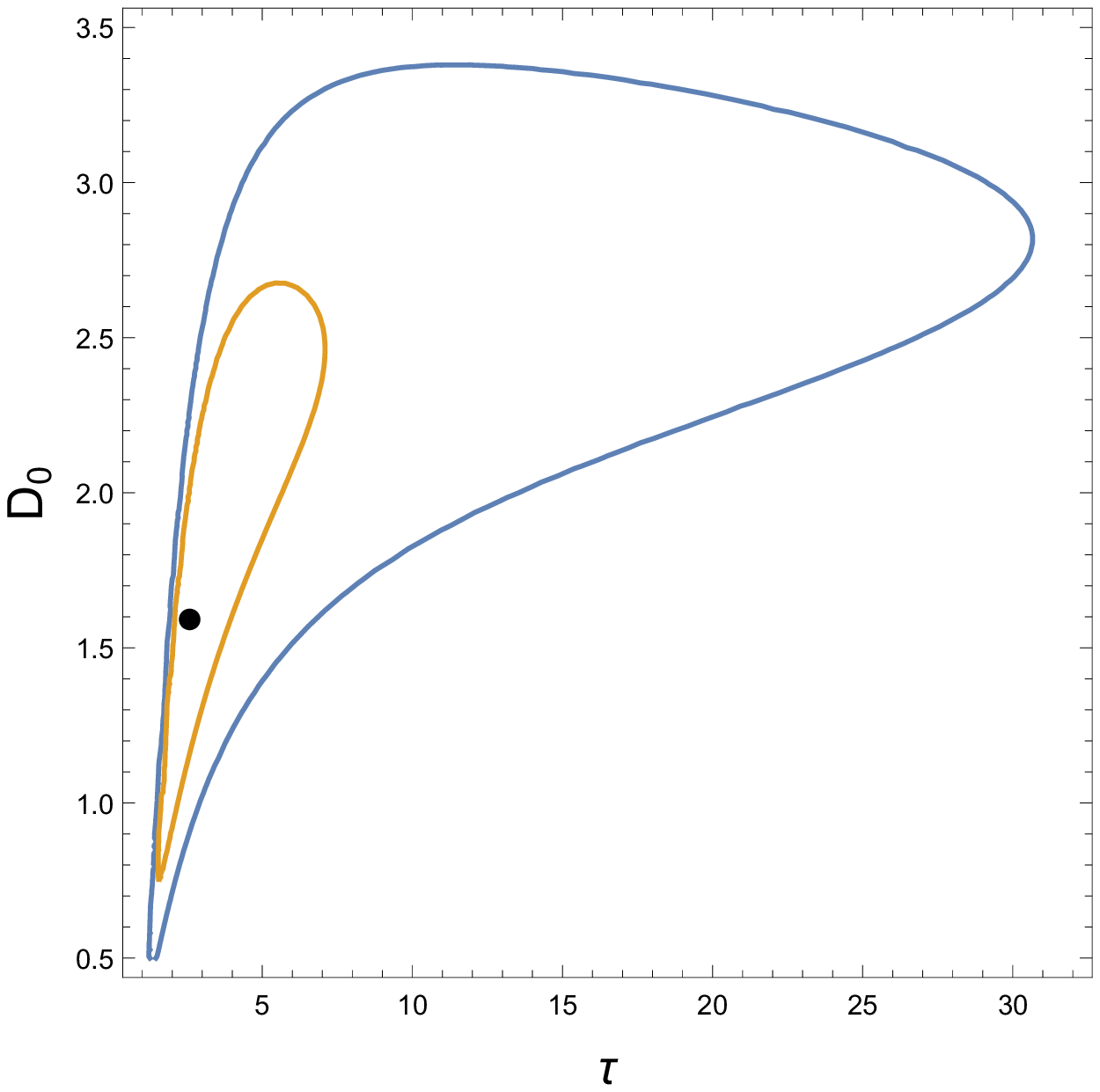}
\end{center}
\caption{The model space of $\tau$ and $\D_0$ (taken to be April 14)
 from the data set for April 15--19. Retrospective changes to the data
 may be possible. The best fit (modal values) is denoted by the black dot.
 The inner curve encloses the \cl{95}, and the outer one the \cl{99}.}
\eef{postset}

\subsection{After April 14--- incomplete data}

We have noted earlier that retrospective changes are expected often to the
most recent fatality data. This is why the analysis of the data for April
15--19 is made separately. Taking day zero to be April 14, we find
\beq
 \tau = 2.6{\rm\ days},\qquad{\rm and}\qquad \D_0 = 1.59. 
\eeq{post}
The joint \cl{95} of the parameters is shown in \fgn{postset}.  The
doubling time is half of that in \eqn{model}, implying an uncontrolled
growth rate for the epidemic. There is less than a 11\% chance that
this doubling rate is compatible with that in \eqn{model}. Even more
disturbing is that $\D_0$ is very low, with a negligible chance (less
than one part in $10^{-30}$) that it is compatible with the lower bound
on the \cl{95} of fatalities allowed by \eqn{model} on April 14. It is
clear from this that the data is either not yet complete, or that we are
confronted with a huge statistical deviation away from the previous part
of the same time series.

\section{Conclusions}

\subsection{The doubling time}

The time series of data on fatalities in Mumbai was used to extract
several characteristics of the epidemic in Mumbai. The most robust
parameter that comes out of this is the doubling time, $\tau=5.3$ days
(median, with $[4.6,6.9]$ \cl{95}). It is surprising that this rate of
growth for COVID-19 is seen during the lock down. It is comparable
to the doubling times reported in the early days of the epidemic in
Wuhan, before a quarantine was imposed there \cite{sfr}.

We also infer a parameter $\D_0$ which is the number of estimated
fatalities on day zero of the analysis, which we take to me March 30. We
find $\D_0=1.8$ (\cl{95}, 1.4--2.4). This is in good agreement with the
reported number of 2 on May 30. We note that this agreement indicates
that the stochasticity in daily numbers is weak. One supporting evidence
is that the cumulative count of fatalities has grown very smoothly over
this period with the same doubling time as found here. This is a direct
check that the assumption of exponential growth remains valid.

\subsection{Estimation of $R_0$}

The doubling time can be matched to the basic reproduction rate, $R_0$,
in different epidemiological models. Here we match it to a simple SEIR
model. This has been used widely to model the COVID-19 epidemic, since
the disease is non-infective in the incubation state. In the early stages
of an epidemic, when the fraction of susceptibles in the population is
close to unity, one may write
\beq
 R_0 = 1 + \log2\,\left(\frac{t_2}\tau\right),
\eeq{sir}
where the distribution of $t_2$ used is given after \eqn{ioest}.
Using the value of $\tau$ quoted in \eqn{model}, we find $R_0=3.4$
(median) and $[2.4,4.8]$ \cl{95}.  This is consistent with estimates from
a range of countries, but closer to the upper edge of the global spread
in values. Note that the only inputs here are the doubling time $\tau$
and the case resolution time $t_2$. There are public health implications
which we discuss later.

\subsection{Counts of positive tested population}

\bef
\begin{center}
\includegraphics[scale=0.5]{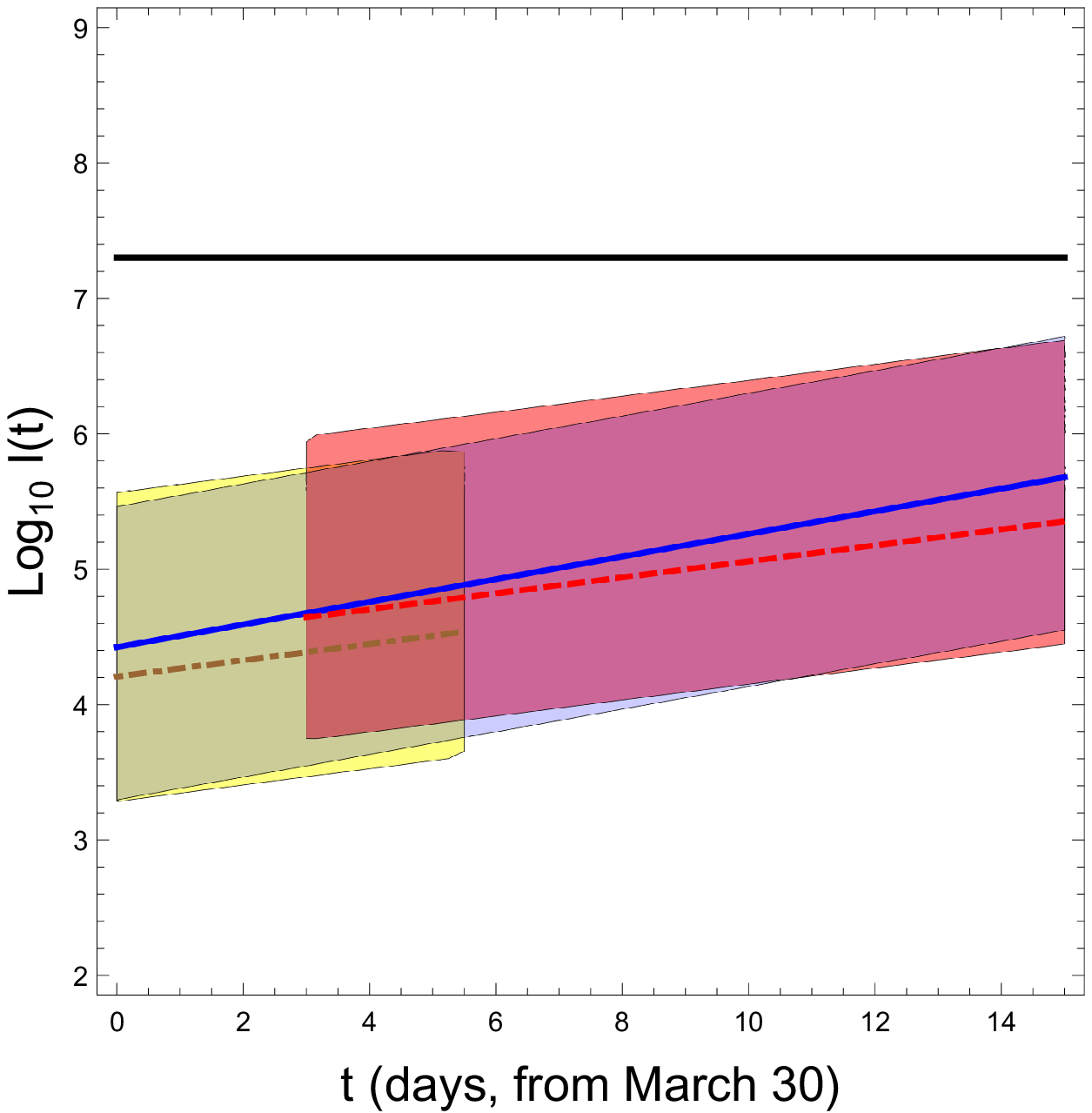}	
\end{center}
\caption{A Bayesian estimate of the number of infections in Mumbai.
 The region in yellow shows the $\cl{95}$ of the estimated number of
 infections inferred from the number of fatalities observed until
 April 4. The brown dash-dotted line shows the median of the distribution.
 The area in pink is the \cl{95} based on data on fatalities from April
 6 onwards, and its median is shown by the red dashed line. Note the
 reasonable agreement in the dates of overlap. The band in blue is
 the \cl{95} for the combined analysis after correcting for improved
 surveillance soon after April 2. The blue line is the median from this
 distribution.  The horizontal black line is the population of Mumbai.}
\eef{infections}

The process of extracting $I_0$ is indirect, since it depends on
three different epidemiological parameters which require studies with
patient databases more extensive than available in Mumbai now. We have
taken them from \cite{ifr}.The \cl{95} shown in
\fgn{infections} should be understood as containing these uncertainties
according to \eqn{ioest}.

MCGM's daily counts of test positives, $T_+(t)$, for Mumbai over the
period of March 31 to April 14 were analyzed with exactly the same methods
which were applied to $D(t)$. It was found that the doubling time, $\tau$,
was 5.7 days (modal value). The modal number of test positives on March
30 was inferred to be $\T_0= 33$.  Given the estimated $I_0$ on that day
\eqn{model}, $\Pi=T_+(t)/I(t)$ can easily be estimated. It was found
that on day zero $\Pi=0.13$\% (median).
Since the doubling time of the numerator and denominator are the same,
this ratio has remained constant through the epidemic.

Without a massive increase in the number of tests, there is no direct
way to verify the estimates of infections presented here. Several alternatives
have been proposed, ranging from testing viral load in
sewage \cite{paris} to conducting sample surveys using different testing
methods. An example of the latter is a recent serological sample study
\cite{seroprev} in California, USA. This study reported $\Pi$ to be
1.1--2\%.  When the test rate, \ie, the number of tests administered per
million of population is small, $\Pi$ is proportional to the test rate.
On April 18, India's cumulative test rates were $2.69\times10^{-4}$,
USA had test rates of $1.1159\times10^{-2}$ \cite{ourworld}. If the
Indian test rate is scaled up to the same value as the US, then, with
the estimates of $\Pi$ presented in the previous paragraph would have to
be multiplied by a factor of 42. This scaled value of $\Pi$ for Mumbai
is compatible with that reported for California in \cite{seroprev}.

\subsection{Dating the beginning of the epidemic}

Given $\tau=5.3$ days, we can extrapolate from $I_0$ backwards to
find when there was only a single case in Mumbai. This estimates the first
seeding of COVID-19 in the city.  We find that the median date for the
initial seeding was around January 15, 2020. However, there is a 95\%
chance that the initial seeding occurred as early as December 20,
2019 or as late as February 1, 2020.  Although the volume of traffic
from China to Mumbai is not large, it is steady, and early seeding is
not inconceivable, especially if the virus was spreading inside China
already in mid-November. The first fatalities due to this cryptic spread
of COVID-19 would not have happened before mid-February (median estimate).
It is certainly possible to check municipal records of deaths in February
to see whether any excesses over the same month in previous years are
visible. If there is a statistically significant signal, it would support
the idea of early cryptic seeding of COVID-19 in Mumbai.

\subsection{Public health implications}

It is interesting that the value of $R_0$ deduced from the growth
rate during a lock down is among the higher end of values seen in
other countries. The purpose of this drastic measure was to reduce
the value of $R_0$. This could be an indication that the epidemic
growth is being driven by high density areas where physical distancing
is impossible. It would be important to understand the structure in
finer detail, ward-wise or at even finer scales, if possible. If there
is indeed such geographical structure to the epidemic within Mumbai,
then it is possible that post-lock down there could be a ``second wave''
surge of cases in areas where housing density is lower. Active steps must
be taken to prevent this. The Maharashtra Government order of April 18,
removing all relaxations of lock down in the Mumbai and Pune regions
\cite{gomo} may be seen in this context.

It should be noted that a significant fraction of the infected could
be asymptomatic \cite{diapr,evacu,cdc}.  In this connection note that
the symptom fatality ratio, $\sfr$, defined as the probability of
dying after the onset of symptoms is reported to be 1.4\% (median)
$[0.9\%,2.1\%]$ \cl{95} \cite{sfr}.  Comparing this to the $\ifr$ of
0.657\% gives an estimate of the probability of being asymptomatic: $P_A=
1-\ifr/\sfr$. This is about 50\%, and in the range quoted in literature.
So it is possible that half the cases are not of significance for public
health measures \cite{cdc,who}. Symptomatic cases are of significance. For
these contact tracing, where possible, and quarantine, otherwise, remain
the best tools available now for controlling spread by bringing down
the growth rate.

However, the asymptomatic fraction is of importance for a long term
solution to the COVID-19 epidemic. Barring a sudden discovery and
large-scale deployment of a cure or vaccine, the epidemic can be only
be stopped when the number of infected reaches a level where herd
immunity begins to operate. Asymptomatic infections have to be counted
in this. From this point of view one should see the upper ends of the
bands in \fgn{infections} as the most optimistic scenarios for reaching
this desirable end state to the epidemic.

Note that if the growth rate is not controlled quickly, the situation
could become much worse. With the estimate of $R_0$ made here, it
should be expected that herd immunity sets in when 71\% (median and
$[58\%,79\%]$ \cl{95}) of the population has been infected. Given
$\ifr=0.657$\%, this could imply around 100,000 fatalities in Greater
Mumbai until herd immunity is reached.  Note that conclusions like this
may be model dependent. This extremely simple model is quoted here only
for orientation.

Another way to look at these predictions is the following. The
fatality rates till now have not been severe. However, at the current
rate of doubling, it would take about 15 days for them to climb about
10-fold. This increase would give about 100 fatalities a day, and could
climb above that. This is in the range of the peak number of daily
COVID-19 fatalities in New York City \cite{nychealth}.

If the fatality rate increases to such levels, the number of cases
requiring critical care increases even more. Also critical care lasts
longer for cases which recover. In the hypothetical cases discussed
here, COVID-19 wards in the city would have to deal with several hundred
admissions a day with care periods for each case ranging up to two months.
Again, using the estimates given above, these rates would be reached
in two to four weeks if the rate of growth of the epidemic does not
come down.

\section{Acknowledgements}

I would like to thank Basudeb Dasgupta, Sandeep Krishna, S.\ Krishnaswamy,
Rajdeep Sensarma, and R.\ Shankar for careful readings of the manuscripts
and useful suggestions, and the ISRC (Indian Scientists' Response to
COVID-19) mailing list for generating questions.


\begin{thebibliography}{99}
\bibitem{nychealth}
 New York City Health Services website
 [https://www1.nyc.gov/site/doh/covid/covid-19-data.page]
\bibitem{reuters}
 Maurice Tamman, 
 {\sl At-home COVID-19 deaths may be significantly undercounted in New York City\/},
 Reuters, April 8, 2020.
 [https://in.reuters.com/article/health-coronavirus-fdny/at-home-covid-19-deaths-may-be-significantly-undercounted-in-new-york-city-idINKBN21Q0E4]
\bibitem{forbes}
 A.\ Galeotti, P.\ Surico, S.\ Hohmann,
 {\sl How many people really die from COVID-19? Lessons from Italy\/},
 Forbes, April 6, 2020.
 [https://www.forbes.com/sites/lbsbusinessstrategyreview/2020/04/06/how-many-people-have-really-died-from-covid-19/]
\bibitem{ifr}
 R.\ Verity, \etal,
 {\sl Estimates of the severity of Coronavirus disease 2019: a model based analysis\/},
 Lancet Infectious Diseases, (2020)
 [https://doi.org/10.1016/S1473-3099(20)30243-7]
\bibitem{seroprev}
 E.\ Bendavid \etal,
 {\sl COVID-19 Antibody Seroprevalence in Santa Clara County, California\/},
 medRxiv preprint April 11, 2020.
 [https://doi.org/10.1101/2020.04.14.20062463].
\bibitem{yadav}
  P.\ D.\ Yadav, V.\ A.\ Potdar, M.\ L.\ Choudhary, \etal,
  {\sl Full-genome sequences of the first two SARS-CoV-2 viruses from India\/},
  Indian J Med Res [Epub ahead of print] [cited 2020 Apr 22]. 
  [http://www.ijmr.org.in/preprintarticle.asp?id=281471]
\bibitem{prahlad}
 P.\ Harsha, data set available at
 https://docs.google.com/spreadsheets/d/1-OYukZzMlRcRKfMqh-pAle0JYAjLjUy08N9V6S51sq0/
\bibitem{lahiri}
 S.\ Gupta and A.\ Lahiri,
 {\sl Ground state \/},
 Eur.\ Phys.\ Lett., 128, 1, 11003 (2019); eprint arxiv:1910.11384.
 [doi:10.1209/0295-5075/128/11003]
\bibitem{previous}
 S.\ Gupta and R.\ Shankar, 
 {\sl Estimating the number of COVID-19 infections in Indian hot-spots using fatality data\/},
 eprint arxiv:2004.04025
 [https://arxiv.org/pdf/2004.04025]
\bibitem{russell}
 T.\ W.\ Russell, J.\ Hellewell, K.\ van Zandvoort, \etal,
 {\sl Estimating the infection and case fatality ratio for coronavirus disease (COVID-19) using age-adjusted data from the outbreak on the Diamond Princess cruise ship, February 2020\/}
 Euro Surveill, 2020 Mar 25(12)
 [doi: 10.2807/1560-7917.ES.2020.25.12.2000256]
\bibitem{cebm}
  J.\ Oke and C.\ Heneghan,
 {\sl Global Covid-19 Case Fatality Rates\/},
  [https:/https://www.cebm.net/covid-19/global-covid-19-case-fatality-rates//www.cebm.net/covid-19/global-covid-19-case-fatality-rates/]
\bibitem{goli}
  S.\ Goli and K.\ S.\ James,
  {\sl How much of SARS-CoV-2 Infections is India detecting? A model-based estimation\/},
  medRxiv preprint, April 15, 2020.
  [doi: https://doi.org/10.1101/2020.04.09.20059014]
\bibitem{lauer}
 S.\ A.\ Lauer, K.\ H.\ Grantz, Q.\ Bi, \etal,
 Ann. Int. Med, 2020
 [doi:10.7326/M20-0504]
\bibitem{li}
  Q.\ Li, X.\ Guan, P.\ Wu, \etal,
 {\sl Early Transmission Dynamics in Wuhan, China, of Novel Coronavirus–Infected Pneumonia\/},
  N Eng J M ed, 2020; 382:1199-1207
  [DOI: 10.1056/NEJMoa2001316]
\bibitem{news}
 T.\ Barnagarwala,
 {\sl State health officials criticise BMC as cases rise in Mumbai\/},
 Indian Express (Mumbai edition), April 2 (2020) 4.
 [https://epaper.indianexpress.com/2619796/Mumbai/April-03-2020]
\bibitem{sfr}
  J.\ T.\ Wu, K.\ Leung, M.\ Bushman, \etal,
  {\sl Estimating clinical severity of COVID-19 from the transmission dynamics in Wuhan, China\/},
  Nature Medicine 26, 506--510 (2020)
  [https://doi.org/10.1038/s41591-020-0822-7]
\bibitem{paris}
  S.\ Wurtzer, V.\ Marechal, J.-M.\ Mouchei, L.\ Moulin,
  {\sl Time course quantitative detection of SARS-CoV-2 in Parisian wastewaters correlates with COVID-19 confirmed cases\/},
  medRxiv preprint April 17, (2020)
  [doi https://doi.org/10.1101/2020.04.12.20062679]
\bibitem{ourworld}
 Our World in Data, [https://ourworldindata.org/grapher/covid-19-tests-cases-scatter-with-comparisons]
\bibitem{gomo}
  Government of Maharashtra, Order no. DMU/2020/CR.92/DisM-1, April 18 (2020),
  {\sl Amendment to the Consolidated Revised Guidelines on the measures to be
  taken for containment of COVID-19 in the State\/}.
\bibitem{diapr}
 K.\ Mizumoto, K.\ Kagaya, A.\ Zarebski and G.\ Chowell,
 {\sl Estimating the asymptomatic proportion of coronavirus disease 2019 (COVID-19) cases on board the Diamond Princess cruise ship, Yokohama, Japan, 2020\/},
 Euro Surveill. 2020 Mar 12; 25(10): 2000180.
 [doi: 10.2807/1560-7917.ES.2020.25.10.2000180]
\bibitem{evacu}
  H.\ Nishiura, T.\ Kobayashi, A.\ Suzuki, \etal,
  {\sl Estimation of the asymptomatic ratio of novel coronavirus infections (COVID-19)\/},
  Int J Infect Dis Published Online First: 13 March 2020.
  [doi:10.1016/j.ijid.2020.03.020]
\bibitem{cdc}
  US Centers for Disease Control,
  {\sl Interim Clinical Guidance for Management of Patients with Confirmed Coronavirus Disease (COVID-19)\/},
  [https://www.cdc.gov/coronavirus/2019-ncov/hcp/clinical-guidance-management-patients.html]
\bibitem{who}
  WHO {\sl Coronavirus Disease 2019 (COVID-19) Situation Report-73\/},
  April 2, 2020.
  [https://Fwww.who.int/docs/default-source/coronaviruse/situation-reports/20200402-sitrep-73-covid-19.pdf]
\end{thebibliography}
\end{document}